\documentclass[fleqn,10pt]{wlscirep}
\usepackage{amsmath,amsfonts,amssymb}
\usepackage{graphicx}
\usepackage{subfigure}
\usepackage[normalem]{ulem}
\usepackage{braket}
\usepackage{amsthm}

\usepackage{color}

\title{All-optical implementation of collision-based evolutions of open quantum systems}

\author[1,*]{\'Alvaro Cuevas}
\author[1]{Andrea Geraldi}
\author[1,2]{Carlo Liorni}
\author[1]{Lu\'is Diego Bonavena}
\author[3,4]{Antonella De Pasquale}
\author[1]{Fabio Sciarrino}
\author[5,*]{Vittorio Giovannetti}
\author[1,*]{Paolo Mataloni}
\affil[1]{Department of Physics, University of Rome La Sapienza, Piazzale Aldo Moro 5, 00185 Rome, Italy}
\affil[2]{Heinrich-Heine-University, Institute for Theoretical Physics III, Universitätsstrasse 1, 40225 Düsseldorf, Germany}
\affil[3]{Department of Physics, University of Florence, Via G. Sansone 1, I-50019 Sesto Fiorentino, Florence, Italy}
\affil[4]{INFN Sezione di Firenze, Via G. Sansone 1, I-50019 Sesto Fiorentino, Florence, Italy}
\affil[5]{NEST, Scuola Normale Superiore and Istituto Nanoscienze, Piazza dei Cavalieri 7, 56126 Pisa, Italy}

\affil[*]{Corresponding authors. Email: alvaro.cuevas@uniroma1.it (\'A. C); vittorio.giovannetti@sns.it (V. G.); paolo.mataloni@uniroma1.it (P. M.)}

\begin{abstract}
We present a new optical scheme enabling the implementation of highly stable and configurable non-Markovian dynamics. Here one photon qubit can circulate in a multipass bulk geometry consisting of two concatenated Sagnac interferometers to simulate the so called collisional model, where the system interacts at discrete times with a vacuum environment. We show the optical features of our apparatus and three different implementations of it, replicating a pure Markovian scenario and two non-Markovian ones, where we quantify the information backflow by tracking the evolution of the initial entanglement between the system photon and an ancillary one.
\end{abstract}

\begin{document}

\flushbottom
\maketitle

\section{Introduction}

Precise control of quantum states is a crucial requirement for future quantum technologies \cite{qcomp,computation_dots}. Their processing protocols should preserve and distribute microscopic correlations in macroscopic scenarios, where countable quantum systems are subjected to environmental noise. It is essential in this context to understand how much robust are the possible quantum dynamical processes and the best way to control the information permeability between the systems and their environment \cite{markovianity_on_demand,markovianity_control,markovianity_photonics,non_markovianity_dephasing,reviewNM3,applNM1}. 

Quantum dynamical processes do not act merely on the sample system, actually they act in an extended Hilbert space where system and its surrounding environment are in contact \cite{computation_markov, markovianity_environment}. The non-isolated sample system $s$ is called \textit{open quantum system} (OQS), and is characterized by a state $\rho_{s}\in\mathcal{H}_{s}$. Similarly, the environment $e$ is characterized by a state $\rho_{e}\in\mathcal{H}_{e}$. Without loss of generality, one can assume that the extended system $s-e$ that lives in $\mathcal{H}=\mathcal{H}_{s}\otimes\mathcal{H}_{e}$ is closed, then no information can be lost but only distributed inside $\mathcal{H}$ \cite{markovianity_memory,markovianity_witness}.

The dynamics of an OQS are called Markovian if each continuous or discrete section of the total evolution is independent of the previous ones, otherwise they are called non-Markovian \cite{markovian_comparison}. In the quantum scenario, three different approaches are widely used to quantify the degree of non-Markovianity of a process \cite{quantum_markovianity,reviewNM2}. The first method is based on the presence of information back-flow towards the system from the environment, that acts in this case as a reservoir of information \cite{markovianity_measure,flow2,markovianity_information}. In the OQS framework the total system-environment state $\rho_{s,e}\in\mathcal{H}$ evolves according to a quantum process generating a communication link between $\mathcal{H}_{s}$ and $\mathcal{H}_{e}$. Here the strength of the flow of information between system and environment during their interaction can be used to discriminate the level of non-Markovianity of the process. The second approach studies the divisibility of the process in \textit{Completely Positive} (CP) maps, defining the evolution as non-Markovian if this decomposition fails at some time \cite{quantum_markovianity,markovianity_detection}. In cases were this CP divisibility is valid, also master equations can be well defined \cite{master_equation}. The third method, which has been used in this work, studies the evolution of the entanglement between the system and an isolated ancilla and it is strictly related to the two approaches mentioned above. In order to explain it we refer to the next section and to \cite{quantum_markovianity,NM_ent}. 

If the environment is represented by an ensemble of spaces $\mathcal{H}_{e}=\mathcal{H}_{e_{1}}\otimes...\otimes \mathcal{H}_{e_{k}}$, and the system space $\mathcal{H}_{s}$ interacts sequentially with each of them at discrete times, we obtain the so called \textit{collisional model} (CM) \cite{collisional1,collisional2,collisional3,collisional10,markovianity_collisional}. It represents a powerful tool to approximate continuous-time quantum dynamics and to analyze non-Markovian dynamics of OQSs \cite{collisional4,collisional5,collisional6,collisional7,collisional8,collisional9}. Linear optics platforms have been thoroughly analyzed for the implementation of CMs \cite{markovianity_simulator,collisional11}.
A simple and effective implementation has been proposed by some of us \cite{markovianity_stroboscopic}. There the authors consider an initial photon state $\rho_{s}(t_{0})\in\mathcal{H}_{s}$, whose spatial mode collides sequentially with the modes of an environment ensemble, which can be considered as a double space environment $\mathcal{H}_{e}=\mathcal{H}_{e_{1}}\otimes\mathcal{H}_{e_{2}}$ with one subspace always prepared in a certain generic state $\rho_{e_{2}}\in\mathcal{H}_{e_{2}}$ at any step-k of the process. The evolution of $\rho_{s}(t_{k})$ is mainly controlled by the interaction involving the Hilbert spaces $\mathcal{H}_{s}$ and $\mathcal{H}_{e_{1}}$, while the memory effects are due to the inter-environment collisions written in $\mathcal{H}_{e_{1}}$ and $\mathcal{H}_{e_{2}}$, which also produces and effective evolution in $\rho_{e_{1}}(t_{k})\in\mathcal{H}_{e_{1}}$. 
\begin{figure}[h!]
	\centering
\includegraphics[width=0.53\columnwidth]{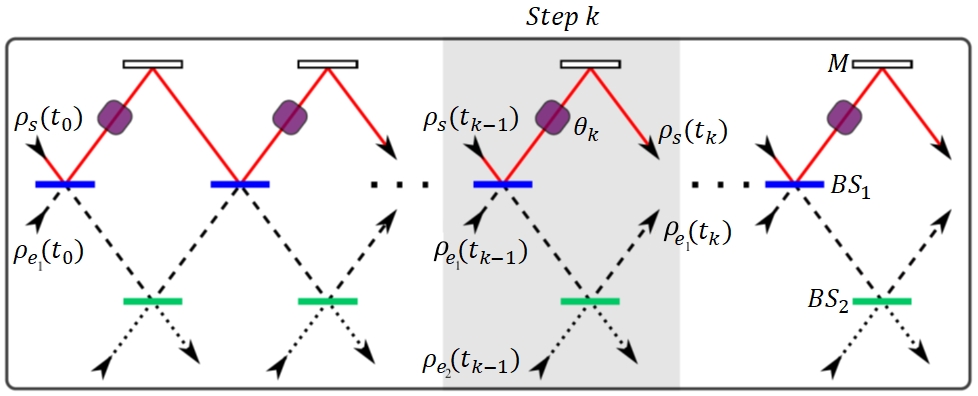}
	\caption{\textbf{Linear optics scheme for the CM:} Each step-k begins with the $s-e_{1}$ collision in $BS_{1}$ (in blue) and it ends with the $e_{1}-e_{2}$ collision in $BS_{2}$ (in green). The phase factor $\theta_{k}$ mediates the collision of the step-(k+1) by optical interference.}
	\label{fig:collisional_1}
\end{figure}  

The optical implementation described in \cite{markovianity_stroboscopic} is realized by a sequence of Mach-Zehnder interferometers (MZIs) as shown in Fig.~\ref{fig:collisional_1}, where the continuous trajectory is associated to $\rho_{s}$, the segmented trajectories to $\rho_{e_{1}}$ and the dotted trajectory to $\rho_{e_{2}}$. At the step-k of this process $\rho_{s}(t_{k-1})$ interferes with $\rho_{e_{1}}(t_{k-1})$ in the \textit{beam splitter} $BS_{1}$, while the inter-environment collision with $\rho_{e_{2}}$ occurs in $BS_{2}$, which posses a variable reflectivity $R_{BS}\in[0,1]$ to control the environment memory.

\section{Theoretical Model}

\begin{figure}[h!]
	\centering
   \includegraphics[width=0.85\columnwidth]{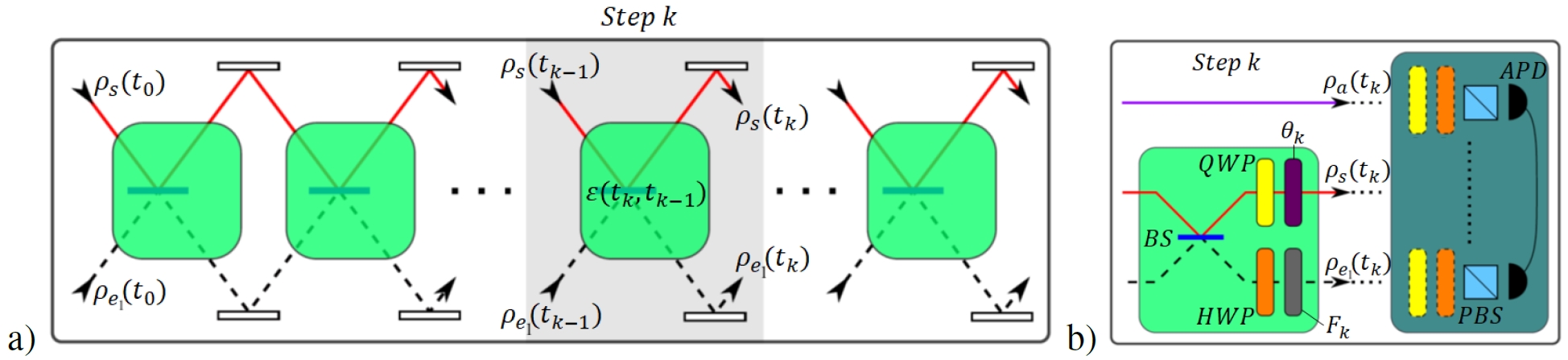}
	\caption{\textbf{Proposed optical scheme, alternative to the one of Fig.~\ref{fig:collisional_1}: a)} When $r_{2}=1$ each $BS_{2}$ is replaced by a perfect mirror $M$. The system-environment collision occurs when $\rho_{s}(t_{k-1})$ interferes with $\rho_{e_{1}}(t_{k-1})$ inside the green box associated to $\epsilon(t_{k},t_{k-1})$. \textbf{b)} The green boxes include a $BS$, a QWP and a phase factor $\theta_{k}$ in the s-mode and a HWP and a filter $F_{k}$ in the $e_{1}$-mode. The environment memory is controlled by $F_{k}$, while $\theta_{k}$ mediates the collision of the step-(k+1) as in the original CM. Two-qubit polarization tomographies can be measured between a-s modes or a-$e_{1}$-modes \cite{state_tomography} by registering the photon coincidences between two avalanche photo-detectors (APDs) after a series of polarization projections through a polarizing beam splitter (PBS).}
	\label{fig:collisional_2}
\end{figure}

In our proposal (shown in Fig.~\ref{fig:collisional_2}a) we consider that all $BS_{2}$ have reflectivity $r_{2}=1$, so that they can be substituted with perfectly reflective mirrors $M$. Here the continuous trajectories correspond to the system ($s$-mode) while the segmented ones correspond to the first environment subspace ($e_{1}$-mode), as in Fig.~\ref{fig:collisional_1}. However, the second environment subspace ($e_{2}$-mode) has no defined path (not present in Fig.~\ref{fig:collisional_2}), since it represents the "absorption environment" after the action of a polarization independent neutral filter $F_{k}$ placed in the $e_{1}$-mode. As seen in Fig.~\ref{fig:collisional_2}b), the super-operator process $\epsilon(t_{k},t_{k-1})$ is composed by a \textit{quarter wave plate} (QWP) in the $s$-mode and a \textit{half wave plate} (HWP) in the $e_{1}$-mode, both at fixed rotation angle $\phi=0$. The environment memory is controlled by the transmissivity factor $T_{k}\in[0,1]$ of $F_{k}$, which gives access to the vacuum state stored by $\rho_{e_{2}}=\ket{0}\bra{0}$, hence effectively mimicking the interaction with the dotted-lines of Fig.~\ref{fig:collisional_1}. The phase factor $\theta_{k}$ mediates collision mechanism by controlling the optical interference. Accordingly, in this setting a purely Markovian dynamics corresponds to the minimum information backflow from $\mathcal{H}_{e_{1}}$ to $\mathcal{H}_{s}$, which is achieved by the maximum loss of information from $\mathcal{H}_{e_{1}}$ to $\mathcal{H}_{e_{2}}$ ($T_{k}=0$). Non-Markovian dynamics instead can arise whenever using $T_{k}\neq0$.
Let us suppose that the s-mode is initially prepared in a maximally entangled state with an external ancillary system (a-mode) as $\ket{\Psi^{\pm}_{a,s}}=\frac{1}{\sqrt{2}}\left(\ket{H}_{a}\ket{V}_{s}\pm\ket{V}_{a}\ket{H}_{s}\right)$, where $\ket{H}$ ($\ket{V})$ represents the horizontal (vertical) polarization of a photon qubit. Since both e-modes are initialized in a vacuum state $\ket{0}$, the actual complete initial state corresponds to $\rho_{a,s,e_{1},e_{2}}(t_{0})=\ket{\Psi^{\pm}}\bra{\Psi^{\pm}}$, with
\begin{equation}
\ket{\Psi^{\pm}}=\frac{1}{\sqrt{2}}
(\hat{a}_{a,H}^{\dagger}\hat{a}_{s,V}^{\dagger}\pm\hat{a}_{a,V}^{\dagger}\hat{a}_{s,H}^{\dagger})\ket{0}_{a,s,e_{1},e_{2}}\equiv\frac{1}{\sqrt{2}}\left(\ket{1_{h}}_{a}\ket{1_{v}}_{s}\pm\ket{1_{v}}_{a}\ket{1_{h}}_{s}\right)\otimes\ket{0}_{e_{1}}\otimes\ket{0}_{e_{2}},
\end{equation}

where $\hat{a}_{x}^{\dagger}$ are the photon creation operators on each x-mode. It is worth stressing that due to the possibility of loosing the s-photon during the propagation after its interaction with the $e_{2}$-mode, our scheme effectively describes the evolution of a qutrit system (with canonical basis given by the states $\ket{1_{h}}_{s}$, $\ket{1_{v}}_{s}$ and $\ket{0}_{s}$), where information is only stored in the bidimensional subspace associated with one-photon sector.

In our prepared scenario the system-environment interactions are controlled by a series of operations, such as the BS one,
\begin{eqnarray}
\hat{BS}_{s,e_{1}}\cdot\left[\ket{1}_{s}\otimes\ket{0}_{e_{1}}\right]&\longrightarrow& i\sqrt{r}\ket{1}_{s}\otimes\ket{0}_{e_{1}}+\sqrt{1-r}\ket{0}_{s}\otimes\ket{1}_{e_{1}},\nonumber\\
\hat{BS}_{s,e_{1}}\cdot\left[\ket{0}_{s}\otimes\ket{1}_{e_{1}}\right]&\longrightarrow& i\sqrt{r}\ket{0}_{s}\otimes\ket{1}_{e_{1}}+\sqrt{1-r}\ket{1}_{s}\otimes\ket{0}_{e_{1}},\\
\hat{BS}_{s,e_{1}}\cdot\left[\ket{0}_{s}\otimes\ket{0}_{e_{1}}\right]&\longrightarrow&\ket{0}_{s}\otimes\ket{0}_{e_{1}},\nonumber
\end{eqnarray}

with $\ket{1}=\left(\alpha\ket{1_{h}}+\beta\ket{1_{v}}\right)/\sqrt{|\alpha|^{2}+|\beta|^{2}}$ and $r$ as its reflectivity factor. The wave plates act according to
\begin{equation}
\hat{HWP}_{s,e_{1}}=\mathbb{I}_{s}\otimes\sigma^{z}_{e_{1}}\quad\text{and}\quad\hat{QWP}_{s,e_{1}}=\sigma^{z/2}_{s}\otimes\mathbb{I}_{e_{1}},
\end{equation}

with $\sigma^{z}=\ket{1_{h}}\bra{1_{h}}-\ket{1_{v}}\bra{1_{v}}+\ket{0}\bra{0}$, $\sigma^{z/2}=\ket{1_{h}}\bra{1_{h}}+i\ket{1_{v}}\bra{1_{v}}+\ket{0}\bra{0}$ and $\mathbb{I}=\ket{1_{h}}\bra{1_{h}}+\ket{1_{v}}\bra{1_{v}}+\ket{0}\bra{0}$.

The attenuation operation applied by the filter $F$ connects the environment space of the remaining light ($\mathcal{H}_{e_{1}}$) with the space of the absorbed light ($\mathcal{H}_{e_{2}}$) according to
\begin{eqnarray}
\hat{F}_{e_{1},e_{2}}\cdot \left[\ket{1}_{e_{1}}\otimes\ket{0}_{e_{2}}\right]&\longrightarrow&\sqrt{T}\ket{1}_{e_{1}}\otimes\ket{0}_{e_{2}}+\sqrt{1-T}\ket{0}_{e_{1}}\otimes\ket{1}_{e_{2}},\nonumber\\
\hat{F}_{e_{1},e_{2}}\cdot \left[\ket{0}_{e_{1}}\otimes\ket{1}_{e_{2}}\right]&\longrightarrow&\ket{0}_{e_{1}}\otimes\ket{1}_{e_{2}},\\
\hat{F}_{e_{1},e_{2}}\cdot \left[\ket{0}_{e_{1}}\otimes\ket{0}_{e_{2}}\right]&\longrightarrow&\ket{0}_{e_{1}}\otimes\ket{0}_{e_{2}},\nonumber
\end{eqnarray}

which generates the effective inter-environment collisions that can reset the $e_{1}$-mode to the vacuum state depending on the absorption factor $1-T$. Finally the phase control acts as
\begin{equation}
\hat{\theta}_{s,e_{1}}=\ket{0}_{s}\bra{0}_{s}\otimes\ket{0}_{e_{1}}\bra{0}_{e_{1}}+\ket{0}_{s}\bra{0}_{s}\otimes\ket{1}_{e_{1}}\bra{1}_{e_{1}}+e^{i\theta}\ket{1}_{s}\bra{1}_{s}\otimes\ket{0}_{e_{1}}\bra{0}_{e_{1}}.
\end{equation}

Then, the super-operator can be written as follows:
\begin{equation}
\epsilon(t_{k},{t_{k-1}})=\mathbb{I}_{s}\otimes\hat{F}_{e_{1},e_{2}}\circ\left(\left(\hat{\theta}_{s,e_{1}}\circ\hat{HWP}_{s,e_{1}}\circ\hat{QWP}_{s,e_{1}}\circ\hat{BS}_{s,e_{1}}\right)\otimes\mathbb{I}_{e_{2}}\right).
\end{equation}

According to our CM represented in Fig.~\ref{fig:collisional_2}, the input state $\rho_{a,s,e_{1},e_{2}}(t_{0})$ evolves as
\begin{equation}
    \rho_{a,s,e_{1},e_{2}}(t_{1},t_{0})=\left(\mathbb{I}_{a}\otimes\epsilon(t_{1},t_{0})\right)\cdot\rho_{a,s,e_{1},e_{2}}(t_{0})\cdot \left(\mathbb{I}_{a}\otimes\epsilon(t_{1},t_{0})\right)^{\dagger},
\end{equation}

 at the first step of the evolution. For consecutive steps, the process can be repeated with variations on $\epsilon(t_{k},t_{k-1})$ or by using the same operation. Finally, one can extract the ancilla-system state as $\rho_{a,s}(t_{k})=Tr_{e_{1},e_{2}}[\rho_{a,s,e_{1},e_{2}}(t_{k})]$ or the ancilla-environment state $\rho_{a,e_{1}}(t_{k})=Tr_{s,e_{2}}[\rho_{a,s,e_{1},e_{2}}(t_{k})]$ by tracing out the undesired spaces and measuring bipartite tomographies after the action of the $k$ single step process.

A characterization of the non-Markovianity of the process can then be obtained by studying the evolution of the concurrence $C_{a,s}$ between the ancilla $a$ and the system $s$ at the various steps of the interferometric propagation. From the results of \cite{quantum_markovianity,markovianity_topology} we know in fact that in the cases where the relation $C_{a,s}(t_{k})>C_{a,s}(t_{k-1})$ holds for some $k>1$, a back-flow of information from $e_{1}$ to $s$ has occurred, resulting in a clear indication of a non-Markovian character of the system dynamics. On the contrary a null increase of $C_{a,s}(t_{k})$ cannot be used as an indication of Markovianity.

The magnitude of all information backflows between two steps of the evolution gauges the degree of non-Markovianity, which can be estimated by considering the integral of the concurrence variation \cite{concurrence,qutrits}, over the time intervals in which it increases, i.e. the quantity 
\begin{equation}\label{eq:measure}
\mathcal{N}=\int_{\dot{C}_{a,s}>0} \dot{C}_{a,s}(t) dt.
\end{equation}

As already mentioned our system $s$ is intrinsically 3-dimensional. Accordingly the $C_{a,s}$ appearing in Eq.~\ref{eq:measure} should be the qutrits concurrence \cite{qutrits} instead of the standard qubit one \cite{concurrence}. However, for the sake of simplicity, in the experimental implementation which we present in the following sections, we shall restrict the analysis only to the entanglement between the single-photon sectors of $s$ and $a$, by property post-selecting our data. Accordingly our measurements do not complete capture the full non-Markovian character implicit in Eq.~\ref{eq:measure}.

\section{Experimental Implementation}

\begin{figure}[h!]
	\centering
 \includegraphics[width=0.53\columnwidth]{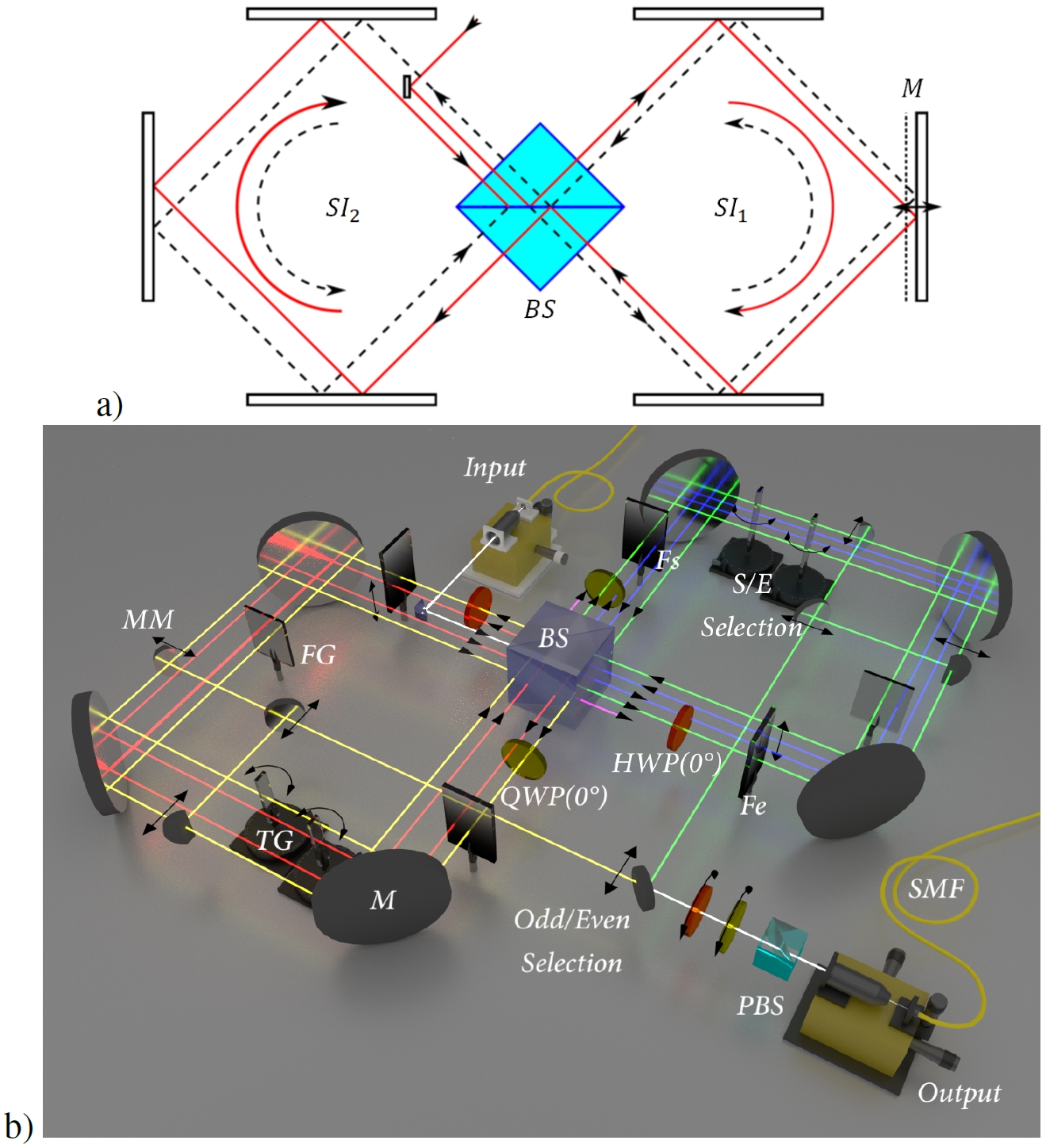}
\caption{\textbf{a) Multipass scheme on a double SI.} The s-modes and e-modes circulating in the clockwise and counter-clockwise trajectories inside each SI, respectively. \textbf{b) Complete setup for the collisional model.} One can extract $\rho_{a,s}(t_{k})$ or $\rho_{a,e}(t_{k})$ by selecting the trajectories direction, the odd or even steps by choosing the SI, and the step number by using the external moving mirrors (MM) with translational stages. We use a single filter $F^{s}$ and a single filter $F^{e}$ for all odd and even steps. The phase factor $\theta_{k}$ is achieved by the tilting glass plate (TG) respect to the fixed glass (FG). Any output qubit can be measured in the tomography stage together with the external ancillary qubit. Here the blue beams correspond to the first step, red beams to second step, the green ones to the third step and yellow ones to the fourth step.}
	\label{fig:setup}
\end{figure}

The experimental setup is based on two concatenated bulk optics Sagnac interferometers (SIs) as described in Fig.~\ref{fig:setup}a). They are initially prepared in a collinear configuration, that by applying the displacement of a mirror in $SI_{1}$ is transformed in a displaced multipass scheme that replicates the CM of Fig.~\ref{fig:collisional_2}. Here we exploit a geometry endowed with high intrinsic phase stability, where different BSs (present in the scheme of Fig.~\ref{fig:collisional_1}) are substituted by different transversal points on a single BS. In this scheme the odd steps circulate in $SI_{1}$, while the even ones circulate in $SI_{2}$. The configuration is equivalent to the model of Fig.~\ref{fig:collisional_2} since we can choose the s-modes and the $e_1$-modes as the clockwise and counter-clockwise trajectories inside each SI, respectively. For the sake of simplicity, from now on we will use the label "e-environment" only for the non absorbed space of the environment, because its complementary part cannot be measured in our configuration.

The relative phase factors $\theta_{k}$ are implemented by a fixed glass plate intersecting all the e-modes inside each SI, while thin glass plates are placed in every s-mode and tilted independently (see Fig.~\ref{fig:setup}b). The transmissivity factors $T_{k}=\frac{T^{e}_{k}}{T^{s}_{k}}$ are implemented by a single neutral density filter $F^{e}$ with transmissivity $T^{e}$ that intersects all the e-modes inside each SI, while another filter $F^{s}$ with transmissivity $T^{s}$ intersects all the s-modes for time-compensation between both optical paths. In this configuration both filters introduce only a controlled absorption, that represents an intrinsic degree of Markovianity under any kind of regime. Even so, the s-e absorptions can be mapped by the relative absorption factor $T_{k}$.
Analogously, a single QWP intersects all s-modes of each SI, while a single HWP intersects all the e-modes. Since s-mode and e-mode contain the same kind of optical elements, we ensure temporally compensated trajectories with an uncertainty of $<30\mu m$ per step. The superposition of the $2^{k}$ trajectories at step $k$ is collected by a single-mode optical fiber (SMF) after the tomography stage of s-e modes. Analogously, another SMF collects the external a-mode.

In this work we focus our attention on the case where all the steps are identical, namely by using a unique filtering factor $T_{k}=F$ and phase factor $\theta_{k}=\theta$. This regime can be described by
\begin{equation}
\epsilon(t_{k},t_{0})=(\epsilon(t_{1},t_{0}))^{k},
\end{equation}

and corresponds to the case of a \textit{stroboscopic evolution} (SE) \cite{markovianity_stroboscopic}.
\begin{figure}[h!]
	\centering
\includegraphics[width=0.8\textwidth]{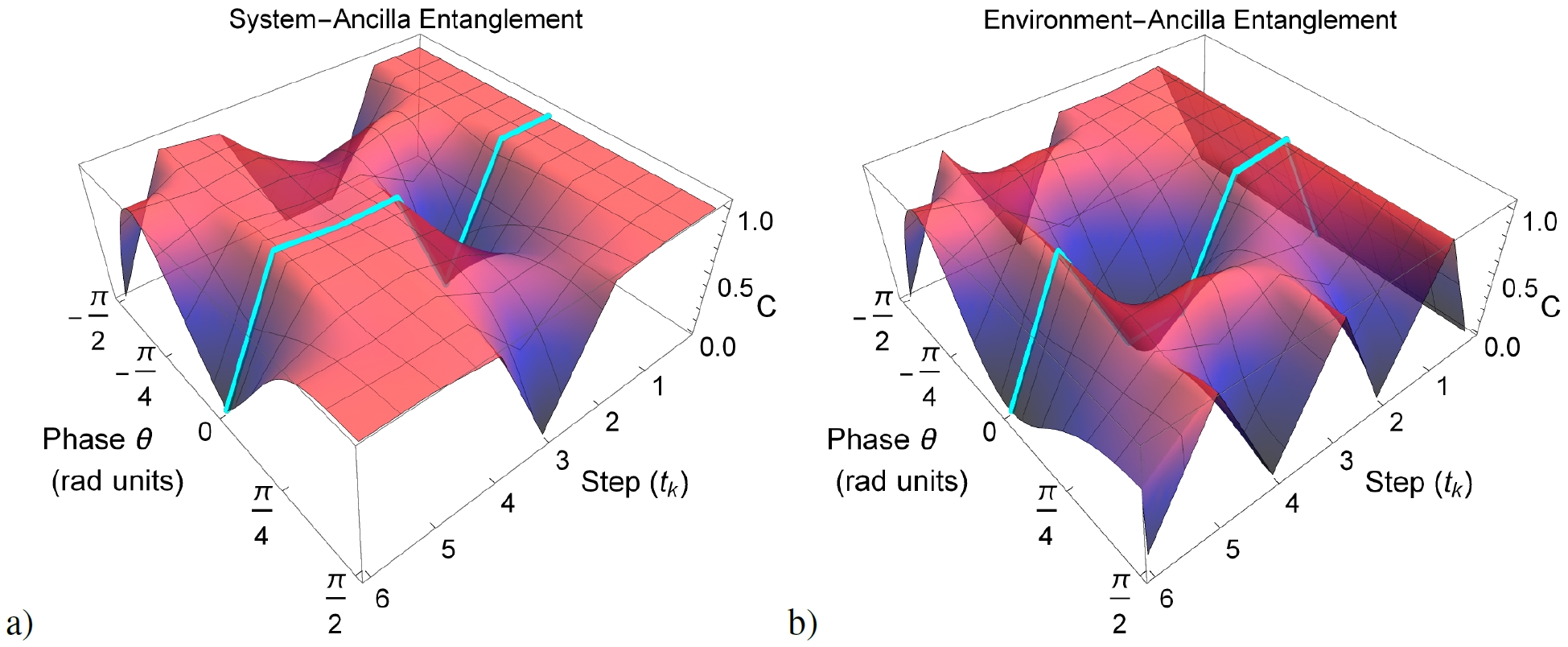}
	\caption{\textbf{Simulated non-Markovian SE of an ideal Bell input state $\ket{\Psi^{\pm}}$ and ideal optics.} Here all the steps are prepared with environment memory parameter $T=1$ (full non-Markovian regime) and phase difference $\theta$. \textbf{a)} Concurrence for system-ancilla state $\rho_{a,s}(t_{k})$. \textbf{b)} Concurrence for environment-ancilla state $\rho_{a,e}(t_{k})$.}
	\label{fig:surfaces}
\end{figure}

The entangled state $\rho_{a,s}(t_{0})=\ket{\Psi^{\pm}_{a,s}}\bra{\Psi^{\pm}_{a,s}}$ is prepared by two indistinguishable processes of Type-II \textit{spontaneous parametric down conversion} (SPDC) inside a high brilliance, high purity Sagnac source based on a periodically-poled KTP (PPKTP) non-linear crystal \cite{source}. Here a single-mode continuous-wave laser at $405nm$ is converted into pairs of photons with orthogonal polarizations at $810nm$ of wavelength and $0.42nm$ of line-width (measured by techniques described in \cite{polariton}). One photon is injected in the s-mode of the setup, while the other travels through the external a-mode. Finally we reconstruct the post-selected state associated to the single-photon sectors of the density matrices $\rho_{a,s}(t_{k})$ or $\rho_{a,e}(t_{k})$ by bipartite hyper-complete tomographies between their associated modes (see Fig.~\ref{fig:collisional_2}b). 

\begin{figure}[h!]
\centering
\includegraphics[width=0.44\textwidth]{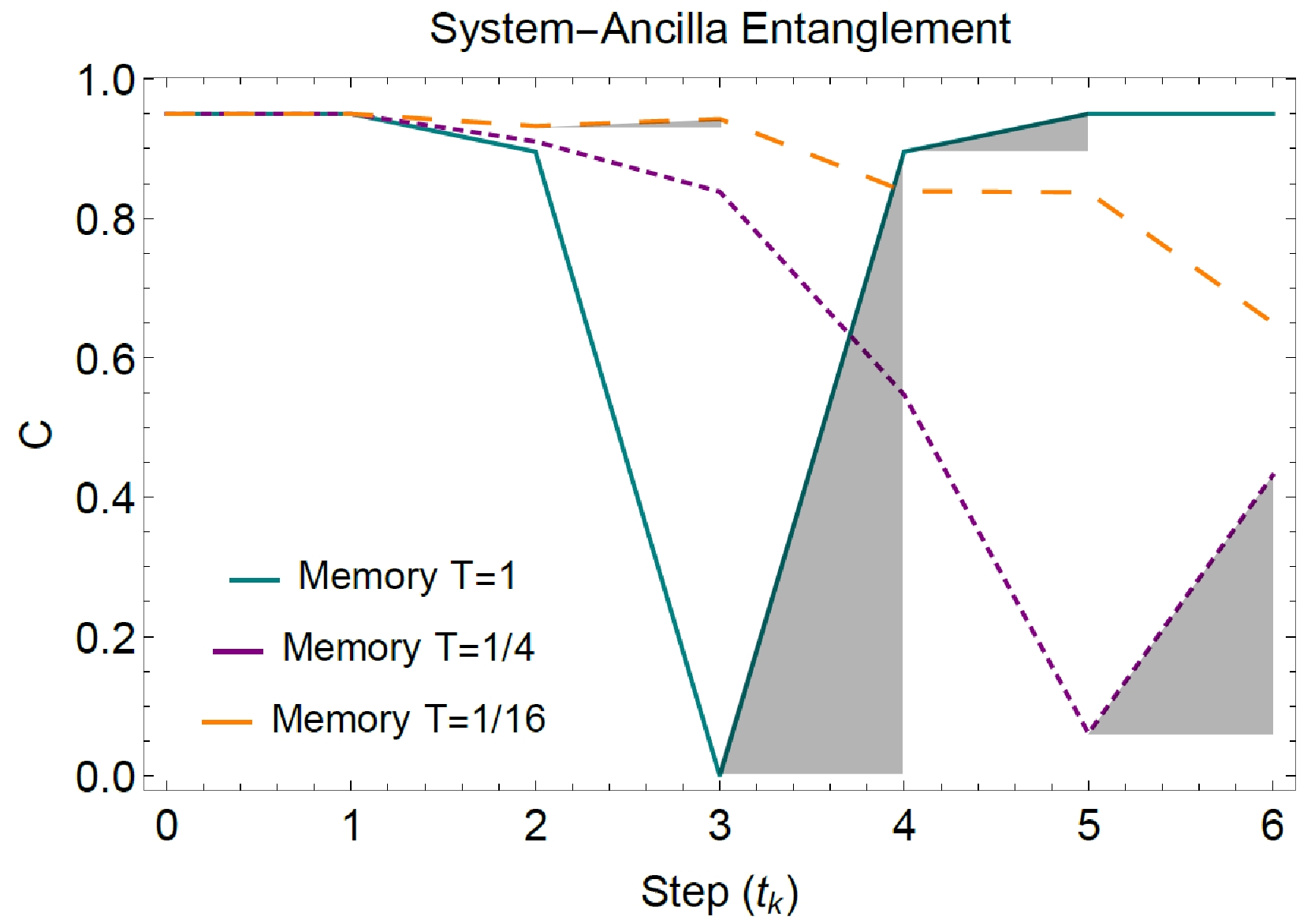}
\caption{\textbf{Simulated non-Markovian SE of the experimental input state $\ket{\Psi^{\pm}}_{exp}$ in the case of ideal optics and $\theta=\pi/4$.} The area of the grey shaded regions correspond to the integral contributions in the quantifier $\mathcal{N}$. For finite number of steps the entanglement revival is lower at lower values of $T$.}
\label{fig:comparison}
\end{figure} 

In Fig.~\ref{fig:surfaces} we show a simulation of the possible non-Markovian dynamics under the SE with maximum environment memory ($T=1$) and variable phase factor $\Phi$. These predicted scenarios were obtained by considering an ideal Bell input state $\ket{\Psi^{\pm}}$ and ideal optical elements, e.g. symmetric BS and no-losses elements. They are interesting to understand and identify the flows and back-flows of information, which can be used in the analysis of engineered $s$-$e$ couplings and its permeability or temporally localized communications for noise avoidance. Besides the s-e collision, the e-mode also suffers inter-environment collisions with the absorption space of the environment. Thus, there is a complex information exchange where it is difficult to identify particular correlations exclusively between both a-s and a-e concurrence behaviours.

In Fig.~\ref{fig:comparison} we show a comparison between three SEs considering ideal optical elements, the actual experimental input state $\ket{\Psi}_{exp}$, a phase factor $\theta=\pi/2$ and different degrees of memory $T$. In the case $T=1$ it results a fast entanglement fluctuation with a non-Markovian degree of $\mathcal{N}=0.475$ up to the sixth step. In the case $T=1/4$ one obtains a slower entanglement fluctuation that gives $\mathcal{N}=0.185$, while in the case $T=1/16$ it emerges an even slower fluctuation with $\mathcal{N}=0.005$ (all values of $\mathcal{N}$ reported here are computed on the post-selected single-photon sectors).

\section{Experimental Results}

The experimental test was restricted to the case of a SE with $\Phi=0$ as seen in the light-blue lines of Fig.~\ref{fig:surfaces}a) and Fig.~\ref{fig:surfaces}b), but considering real optical elements. The prepared entangled state $\Omega_{a,s}$ showed a measured Fidelity $F=|\bra{\Psi_{a,s}^{\pm}}\Omega_{a,s}\ket{\Psi_{a,s}^{\pm}}|=0.9712\pm 0.0004$, then the simulated data for the imperfect evolution referred to a Werner input mixed state \cite{werner,werner_optics,det_qcc} $\Omega_{a,s}=\frac{4F-1}{3}\ket{\Psi^{\pm}_{a,s}}\bra{\Psi^{\pm}_{a,s}}+\frac{1-F}{3}\mathbb{I}_{a}\otimes\mathbb{I}_{s}$.
\begin{figure}[h!]
	\centering
\includegraphics[width=0.85\textwidth]{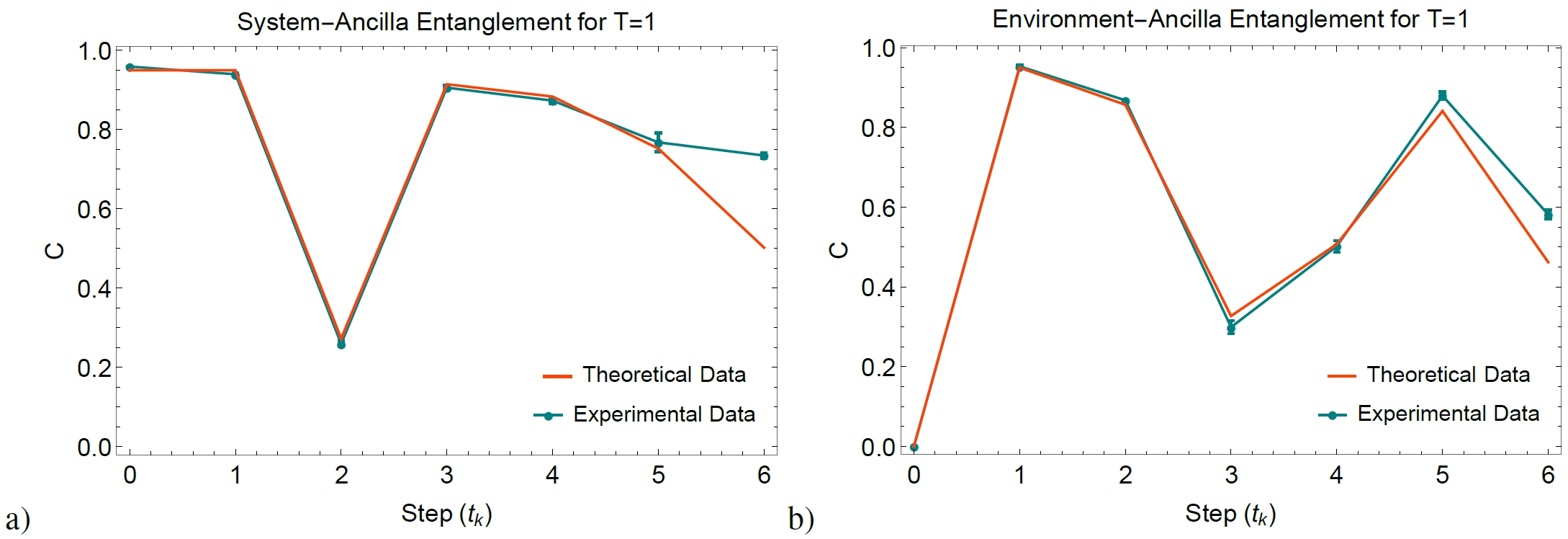}
	\caption{\textbf{Non-Markovian dynamics with maximum memory ($T=1$). a)} Concurrence of $\rho_{a,s}(t_{k})$. \textbf{b)} Concurrence of $\rho_{a,e}(t_{k})$. All error bars were calculated from the propagation of 100 Monte-Carlo simulations with Poisson statistics, while theoretical data were simulated by considering the actual optical elements of the interferometric setup.}
	\label{fig:results_1}
\end{figure}
\begin{figure}[h!]
	\centering
    \includegraphics[width=0.44\textwidth]{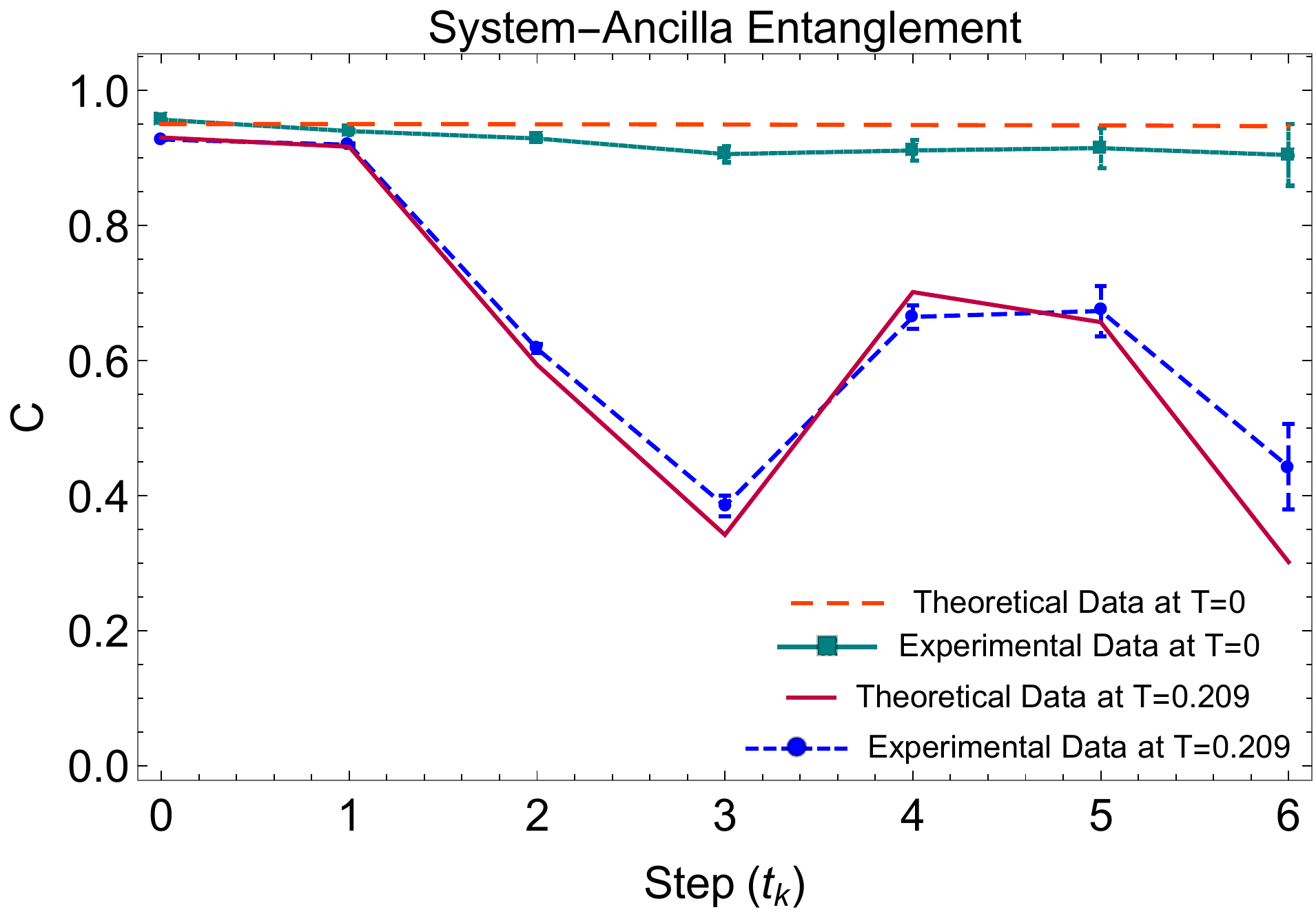}
	\caption{\textbf{Evolution of Polarization Entanglement with reduced memory ($T=0$ and $T=0.209$).} Concurrence of the single-photon sectors, post-select density matrix $\rho_{a,s}(t_{k})$. All error bars were calculated from the propagation of 100 Monte-Carlo simulations with Poisson statistics, while theoretical data were simulated by considering the actual optical elements of the interferometric setup.}
	\label{fig:results_2}
\end{figure}

In Fig.~\ref{fig:results_1} we present the concurrence fluctuations of the single-photon sectors expressed in the post-selected density matrices $\rho_{a,s}(t_{k})$ and $\rho_{a,e}(t_{k})$ during the SE. These states are reconstructed by normalizing the remaining non absorbed coincident photons, and by consequence the associated concurrence values become invariant under losses. Nevertheless, both dynamics behave according to the simulation for the Werner-like input state $\Omega_{a,s}$. In the case $T=1$ of Fig.~\ref{fig:results_1}a) we obtained the highest possible non-Markovianity, where the large concurrence fluctuations give us $\mathcal{N}=0.3232$. As seen in Fig.~\ref{fig:results_2}, in the case $T=0.209$ we obtained reduced concurrence revivals and non-Markovianity of $\mathcal{N}=0.1442$, while in the case $T=0$ we confirmed the lowest possible non-markovianity by obtaining a near to zero value on $\mathcal{N}=0.0044$. 

For our particular CM, these results confirm that entanglement revivals are strictly connected to the environment memory. In fact, they show with high precision that decreasing values on $T$  reduce the information back-flows to the s-mode. The slight deviation from the expected theoretical simulations originates from the not perfect superposition of all possible photon trajectories. Even so, this error is strongly minimized by the use of SMFs as final spatial filters. 

\section{Conclusions}

In this work we presented a linear optics setup that allows to simulate different open quantum systems dynamics. It is based on a novel interferometric structure that guarantees high phase stability and a multipass evolution in a
compact setup, that makes possible to study the dynamics up to 6 steps at least. The dynamics studied here represents the first implementation of the so-called \textit{collisional model} for open quantum systems \cite{markovianity_stroboscopic}, and our results correspond to a particular case of it. The setup is able to simulate a wide variety of stroboscopic evolutions, from strictly Markovian all the way up to strongly non-Markovian dynamics, where quantum memory effects show their contribution. We can experimentally track the role of system-environment and intra-environment interactions in the arising of non-Markovian features and characterize the transition between the two regimes. As the field of quantum technologies spreads, more and more attention has being addressed to the study of non-Markovian dynamics. It can, in principle, be used for efficient information processing \cite{appl1,appl2,violaLloyd,liu2016efficient,dong2018non}, as well as for engineering novel interesting quantum states \cite{appl3,appl4,appl5}. In this perspective, our scheme can be of great interest, thanks to its stability, modular nature and direct access to the environmental degrees of freedom.

\bibliography{References}

\section*{Acknowledgements}

We acknowledge support from the European Commission grants FP7-ICT-2011-9-600838 (QWAD - Quantum Waveguides Application and Development) and H2020-FETPROACT-2014 (QUCHIP - Quantum Simulation on a Photonic Chip). We thank partial support from the Chilean agency Comisi\'on Nacional de Investigaci\'on Cient\'ifica y Tecnol\'ogica (CONICYT) and its Ph.D. scholarships “Becas Chile”.

\section*{Author Contributions}

V. G. and P. M proposed the theoretical frame and the optical scheme presented in Fig. \ref{fig:collisional_2}, \'A. C. proposed and coordinated the experimental multipass implementation, A. G., C. L. and L. D. B. achieved and analysed the experimental measures, A. D. P. and F. S. contributed to the interpretation of results. All authors contributed to the writing of the manuscript.

\section*{Additional Information}

\textbf{Competing interests:} The authors declare that they have no competing interests.

\end{document}